\documentclass[conference]{newIEEEtran}
\usepackage{amsmath}
\usepackage{amssymb}
\usepackage{graphicx}
\usepackage{subfigure}
\usepackage{cite,mcite}
\usepackage{xspace}
\usepackage{url}

\def\EeV{\ifmmode {\mathrm{\ Ee\kern -0.1em V}}\else
                   \textrm{Ee\kern -0.1em V}\fi}%
\def\PeV{\ifmmode {\mathrm{\ Pe\kern -0.1em V}}\else
                   \textrm{Pe\kern -0.1em V}\fi}%
\def\TeV{\ifmmode {\mathrm{\ Te\kern -0.1em V}}\else
                   \textrm{Te\kern -0.1em V}\fi}%
\def\GeV{\ifmmode {\mathrm{\ Ge\kern -0.1em V}}\else
                   \textrm{Ge\kern -0.1em V}\fi}%
\def\MeV{\ifmmode {\mathrm{\ Me\kern -0.1em V}}\else
                   \textrm{Me\kern -0.1em V}\fi}%
\def\eV{\ifmmode {\mathrm{\ e\kern -0.1em V}}\else
                   \textrm{e\kern -0.1em V}\fi}%
\newcommand{\Xmax}{\ensuremath X_\mathrm{max}\xspace}
\newcommand{\meanXmax}{\ensuremath \langle X_\mathrm{max}\rangle\xspace}

\begin{document}

\title{Cosmic Rays above the Knee}

\author{\IEEEauthorblockN{Michael Unger\authorrefmark{1}}
\\
\IEEEauthorblockA{\IEEEauthorrefmark{1}Karlsruher Institut f\"ur Technologie (KIT)\\
Postfach 3640, D-76021 Karlsruhe, Germany\\ Email: Michael.Unger@kit.edu}
}



\maketitle

\begin{abstract}
An overview on the present observational status 
and phenomenological understanding of cosmic
rays above 10$^{16}$~\eV{} is given. Above these
energies the
cosmic ray flux is expected to be gradually dominated by an
extra-galactic component. In order to investigate the nature of this
transition, current experimental activities focus on the
measurement of the cosmic ray flux and composition at the 'ankle' or
'dip' feature at several~\EeV.  At the ultra high energy end of the
spectrum, the flux suppression above 50~\EeV{} is now well established
by the measurements of HiRes and the Pierre Auger Observatory and
we may enter the era of charged particle astronomy.
\end{abstract}

\IEEEpeerreviewmaketitle

\section{Introduction}
\bstctlcite{IEEEexample:BSTcontrol}
 
\begin{figure}
  \includegraphics[width=\linewidth]{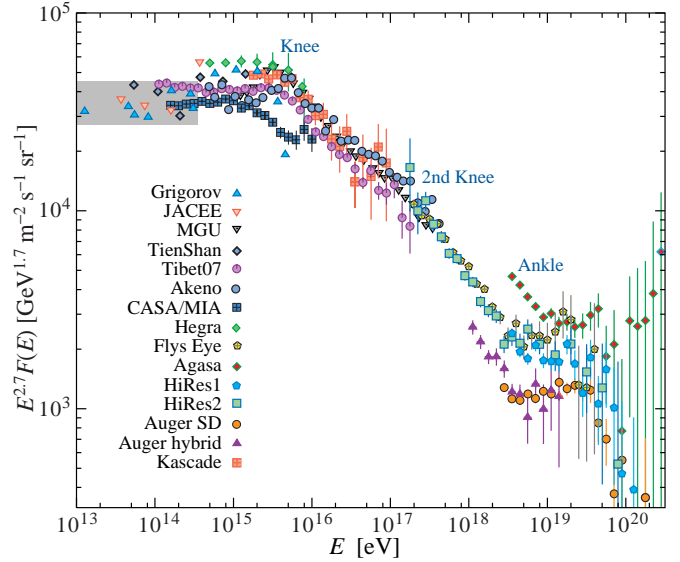}
  \caption[flux]{All particle flux of cosmic rays (\cite{Amsler:2008zz}
   and references therein)}
  \label{fig_pdg}
\end{figure}

The all particle spectrum of cosmic rays is known to follow a power
law, $dN/dE\varpropto E^{-\gamma}$ over many orders of magnitude.
However, at the highest energies, shown in Fig.~\ref{fig_pdg}, it
exhibits three remarkable features.  The {\itshape
  knee}~\cite{kulikov1958}, a steepening of the flux by
$\Delta\gamma\approx 0.5$, at a few~\PeV{} followed by a flattening
called the {\itshape ankle}~\cite{linsley1963} at several~\EeV{} 
and a flux suppression at ultra high energies~\cite{Greisen:1966jv,*Zatsepin:1966jv}.

The first two features are suspected to be an indication
of the end of the galactic cosmic ray spectrum and the transition
to an extra-galactic component.

At energies above several hundreds of \TeV{} the particle fluxes
are too low to allow for a direct measurement of
the properties of cosmic rays.  Instead, as will be explained in
Sec.~\ref{sec:airshowers}, the analysis of {\itshape air showers}
plays a crucial role to measure their flux and composition.  
These two observables are essential to study the
transition from galactic to extra-galactic cosmic rays and to
distinguish between the various models put forward to explain the
ankle (see Sec.\ref{sec:galegal} and~\ref{sec:compo}).

At the same hour this talk was given, the first proton beams
were injected to the Large Hadron Collider~\cite{cernbeams}, that will
eventually be able to accelerate protons up to $7\cdot10^{12}$~\eV.
The ultra-high energy frontier of physics is however beyond
$10^{20}$~\eV, where several cosmic rays have already been
detected\cite{Linsley:1963km,Bird:1994uy,Matthews:2005ve}.  At these
extreme energies, particles are expected to suffer significant energy
losses during their propagation to earth. The corresponding flux
suppression was predicted over forty years
ago~\cite{Greisen:1966jv,*Zatsepin:1966jv} and it is only now, that
experiments gathered enough statistics to study it carefully (see
Sec.~\ref{sec:gzk}).

The astrophysical sources that are able to accelerate particles to
such tremendous energies are still unknown.  Their unambiguous
identification requires to study the arrival directions of cosmic rays,
i.e. to do {\itshape particle astronomy}.

Whereas the experimental knowledge on cosmic rays made a major leap
forward in the current {\itshape hybrid era}~\cite{Watson:2008wv}, new
projects aim to accumulate more data below 1~\EeV{} and at ultra high
energies.  Moreover, new data on hadronic interactions at man-made
particle accelerators are needed to facilitate the interpretation of
air shower data~\cite{Engel:2002id}.  These efforts will be described
in Sec.~\ref{sec:outlook}.
\section{Air showers}
\label{sec:airshowers}
Cosmic particles entering the earth's atmosphere sooner or later
collide with the nuclei of the air and initiate a particle cascade,
the so-called air shower. Since the thickness of the atmosphere is more
than 20~radiation and interaction lengths at vertical incident, it
constitutes a suitable calorimeter to study the properties of the
primary cosmic ray particles.\\ Air shower detectors either measure
the lateral densities of particles at ground or the longitudinal
development of the cascade in the atmosphere.

The qualitative relation of these experimental observables to the
energy $E_0$ and mass $A$ of the primary particle can be easily
understood within the simple Heitler-model~\cite{CarlsonOppenheimer,
  Heitler, Matthews_HeitlerModel,Horandel:2006jd} of air showers. Here
one assumes that after each hadronic interaction length, $\lambda$,
$\pi$-mesons are produced with an average multiplicity of $\langle n
\rangle$. In each interaction an energy fraction of $f\approx 1/3$
goes to neutral pions which decay immediately into two photons and
thus feed the electromagnetic component of an air shower, that
develops through pair-production and bremsstrahlung until the energy
of electromagnetic secondaries falls below the critical energy
$\epsilon_\mathrm{em}$ ($\approx 81$~\MeV{} in air).  In that way, the
fraction of the primary energy in the electromagnetic component at the
$n$th interaction at depth $n\cdot\lambda$ increases to $1-(1-f)^n$
until the energy of the charged pions falls below the critical energy
$\epsilon_\mathrm{ch}$ at which their decay length becomes smaller
than the interaction length. If one furthermore assumes that a nucleus
with mass $A$ and energy $E_0$ is equivalent to $A$ nucleons of energy
$E_0/A$ (the so-called {\itshape superposition model}), the following
important relations can be deduced from this simplistic model:

The average depth at which the electromagnetic cascade reaches its
maximum, $\meanXmax$, grows logarithmically with the energy per nucleon:
\begin{equation}
  \meanXmax = a + b\lg\left[(E_0/\epsilon_\mathrm{em})/A\right],
  \label{eq:xmax}
\end{equation}
where the constants $a$ and $b$ depend on the properties of hadronic
interactions, $f$, $\lambda$ and $\langle n\rangle$. Since most of the
energy of the primary particle eventually ends up in the
electromagnetic cascade, the integral electron number is a good 
estimator for $E_0$. Hence, the observation of the longitudinal development of
an air shower with for instance fluorescence
detectors~\cite{Porter:1970et}, allows to measure simultaneously the
primary energy and $\Xmax$ and can therefore be used to determine the
absolute value of the average nuclear mass as a function of energy
(cf. Sec.~\ref{sec:compo}). The {\itshape elongation
  rate}~\cite{Linsley1977, Gaisser1979, Linsley:1981gh},
$\mathrm{d}\meanXmax/\mathrm{d}(\lg E)\varpropto - \mathrm{d}(\lg
A)/\mathrm{d}(\lg E)$, can be used to study the change of the primary cosmic ray
composition with energy.

The number of muons, $N_\mu$, from the decay of charged pi-mesons 
that can be detected by particle detectors on ground is given by
\begin{equation}
   N_\mu = \left(E_0/\epsilon_\mathrm{ch}\right)^\beta A^{1-\beta}
  \label{eq:sizevsE}
\end{equation}
where again the properties of hadronic interactions are hidden in a
single number, $\beta$, that is proportional to the logarithm of the
charged meson multiplicity. Given a similar relation for the number of
electrons on ground, a detector that is capable to distinguish muons
and electrons can therefore disentangle the energy and mass of cosmic
primaries on a statistical basis.\\ 

The simple Heitler approach
is very useful to understand the principles of air shower physics,
 but of course in practice experiments employ full Monte Carlo simulations 
of air showers with for instance \texttt{CORSIKA}~\cite{Heck:1998vt}
to interpret their data. These simulations are however of limited
predictive power, as they need to rely on models of hadronic
interactions at energies beyond man-made accelerators. The
related uncertainties are the source of considerable systematic uncertainties
for the interpretation of air shower data (see \cite{Knapp:2002vs, 
Pierog:2006qu}).

\begin{figure}
  \includegraphics[width=\linewidth]{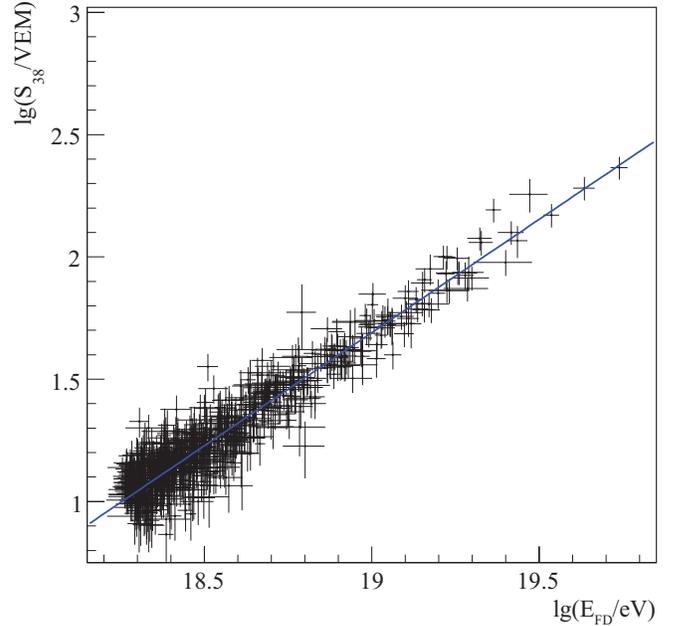}
  \caption[augerCalib]{Energy calibration of the surface detector
                       of the Pierre Auger Observatory~\cite{Abraham:2008ru}: 
                       Shower size, $S_{38}$, as a function of energy, 
                       $E_\mathrm{FD}$, measured
                       with the fluorescence detector.}
  \label{fig_augercal}
\end{figure}

Since the energy estimated from the integral of the electromagnetic 
longitudinal air shower development depends only very 
little on details of hadronic 
interactions~\cite{Barbosa:2003dc,Pierog:2007zza}, modern air
shower arrays like the Pierre Auger Observatory~\cite{Abraham:2004dt} or 
Telescope Array~\cite{Kawai:2008zza} use a {\itshape hybrid} approach to
calibrate the energy scale of their surface detector with the energy estimate
of a fluorescence detector. An example is shown in Fig.~\ref{fig_augercal},
where the expected power-law dependence of the number of ground particles on
the primary energy (Eq.~(\ref{eq:sizevsE})) can be seen.
\section{From Galactic to Extra-galactic Cosmic Rays}
\label{sec:galegal}
The standard explanation for the knee-feature in the cosmic ray
all-particle flux between $10^{15}$ and $10^{16}$~\eV{} is that it
marks the beginning of the end of the galactic cosmic ray spectrum due
to the escape of the high energy charged particles from the magnetic
confinement within the galaxy and/or the reach of the maximum energy
of galactic accelerators (presumably supernova remnants). Although the
current experimental data can not rule out alternative explanations
for the knee (eg.~\cite{Nikolsky:1995gw,Erlykin:1997bs, Dova:2001nd,
  Candia:2002eg, Dar:2006dy}), extrapolations of the low energy cosmic
ray data with a rigidity dependent cut off $\varpropto Z\cdot
E_\mathrm{c}$ can describe the existing data very
well~\cite{Hoerandel:2002yg, Berezhko:2007gh}. Moreover, the
deconvoluted galactic mass spectra measured with
KASCADE~\cite{Antoni:2005wq} show distinct knees for each elemental
component, compatible with a rigidity dependent knee.\\ Since
$E_\mathrm{c}$ is of the order of \PeV{}, it follows that above
energies of several $10^{18}$~\eV{}, the detected cosmic particles
must be of extra-galactic origin. As a corollary, this assumption
explains the lack of an observation of a strong anisotropy that would
be expected for charged particles with a large gyro-radius at this
energy.

 In the following we will describe three different
models of the transition from galactic to extra-galactic cosmic rays 
(see~\cite{Hillas:2006ms,Stanev:2006ri,Berezinsky:2007wf, DeDonato:2008wq} for
 recent reviews on this topic). Common to all of these models is
that the properties of extra-galactic cosmic ray sources are described 
by just four parameters: the source emissivity (needed to adjust
the overall normalization), the spectral index $\gamma_0$ of the energy
spectrum at the source, the maximum energy $Z\cdot E_\mathrm{c}$ the source
is able to accelerate particles to and the cosmological source evolution 
parameter $m$, that describes the source density $n$ as a function of redshift
$z$, $\mathrm{dn}/\mathrm{dz}\varpropto (1+z)^m$. In order
to simplify the discussion, we will restrict ourselves to the
uniform source distribution model ($m=0$) and assume that $E_\mathrm{c}$
is large enough to have no observational consequence within the statistical
precision of current experiments. Furthermore we will assume the simple
phenomenological rigidity-dependent parameterization 
from~\cite{Hoerandel:2002yg} to describe the 'standard' galactic cosmic
ray component. Since there is an obvious disagreement of the measured
ultra-high energy spectra~\cite{Takeda:2002at,Abbasi:2007sv,Yamamoto:2007xj}
(cf.~Fig~\ref{fig_pdg}), their energy scale will be 'adjusted' accordingly
within the quoted systematic uncertainties.
\subsection{Dip Model}
In the so-called {\itshape dip model}~\cite{Berezinsky:1987kj,*Berezinsky:2002nc}, 
all extragalactic cosmic rays are assumed to be protons (at
least after they escaped from their sources~\cite{Sigl:2005md})
and the ankle feature is caused by energy losses
suffered during the propagation to earth.

\begin{figure}
\begin{center}
\subfigure[redshift]{\includegraphics[width=0.86\linewidth]{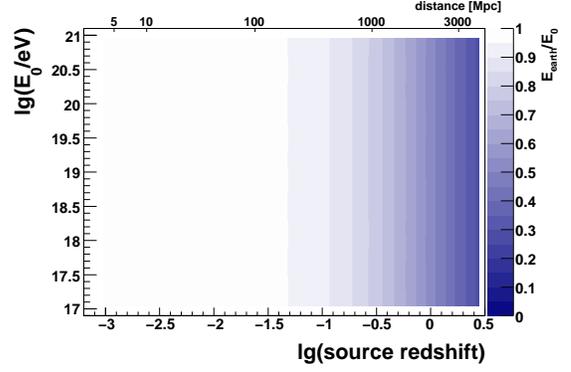}
\label{fig_loss1}}
\subfigure[redshift and pair-production]{\includegraphics[width=0.86\linewidth]{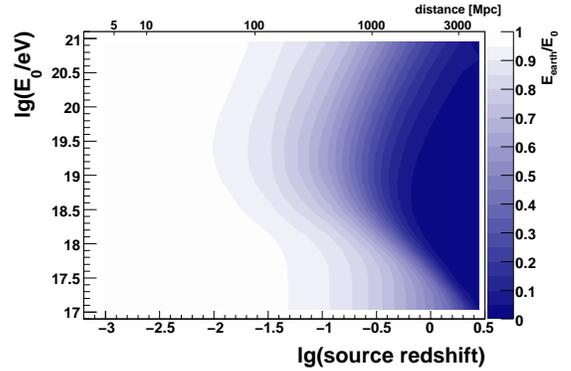}
\label{fig_loss2}}
\subfigure[redshift, pair- and photo-pion-production]{\includegraphics[width=0.86\linewidth]{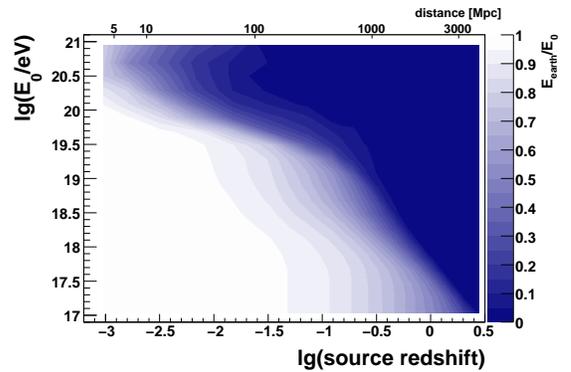}
\label{fig_loss3}}
\end{center}
\caption[energy loss]{Fractional energy at earth, $(E_\mathrm{earth}/E_0)$,
  of protons with initial energy $E_0$ 
    as a function of the source distance/redshift
   ({\scshape CRpropa}~\cite{Armengaud:2006fx}
  calculation). }
\label{fig_protonLoss}
\end{figure}

First of all, the expansion of the universe causes adiabatic losses
that are important for very distant sources 
(cf.~Fig.~\ref{fig_protonLoss}a). Since it affects all energies equally,
it does not change the spectral index observed at earth.\\
A more important effect is the interaction of the cosmic ray protons 
with the photons of the cosmic microwave background radiation.
At lower energies, the production of $\mathrm{e}^+\mathrm{e}^-$-pairs
through the Bethe-Heitler process is the dominant source of energy loss 
(cf.~Fig.~\ref{fig_protonLoss}b) and at energies above $10^{19.5}$~\eV{},
the photon-proton center of mass energy is large enough for resonant
photo-pion production, that gives rise to
large energy losses~\cite{Greisen:1966jv,*Zatsepin:1966jv} even
for very close source (cf.~Fig.~\ref{fig_protonLoss}c).

As can be seen in Fig.~\ref{fig_flux}a, the dip model can describe
the data rather well. The transition between the 
galactic and extra-galactic cosmic ray component is at low
energies just above 0.1~\EeV{} and produces the somewhat less 
prominent feature called the {\itshape second knee}~\cite{Bergman:2007kn}.\\
Thus the dip model is a very economic approach in terms of assumptions,
as it can explain all features in the cosmic ray energy spectrum 
in terms of the well understood interactions of protons with photons
and predicts a transition energy that is low enough
to be compatible with current estimates of the maximum 
energy of galactic accelerators~\cite{Berezhko:1999sc}.\\
However, the hard injection spectrum at the source
is problematic in terms of the overall energy luminosity of extra-galactic
sources if extrapolated to low energies. Therefore
an 'artificial' softening of the spectrum below a
certain energy is usually introduced~\cite{Berezinsky:2004wx} 
(not shown in Fig.~\ref{fig_flux}a).  
\subsection{Mixed Composition Model}
The dip-model works
only for a pure proton beam since an admixture of 
heavier nuclei with a fraction of $\ge 15\%$ diminishes the agreement 
with the data considerably. This is, because the threshold for  
$\mathrm{e}^+\mathrm{e}^-$-production is proportional to the energy
per nucleon and thus only relevant for protons at the
energies of the ankle. Instead,
cosmic ray nuclei loose their energy predominantly 
due to photo-disintegration at the giant dipole
resonance~\cite{Khan:2004nd}. The mean free path for photo-disintegration 
scales with the Lorentz-factor of the particle and drops rapidly
above $\Gamma\gtrsim 10^9$.

\begin{figure}
\begin{center}
\subfigure[Extragalactic protons]{\includegraphics[width=0.81\linewidth]{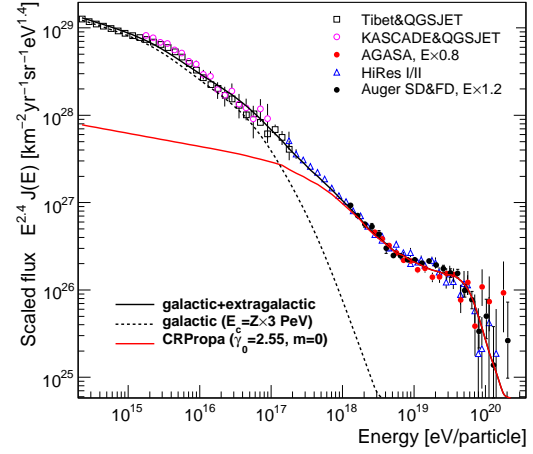}
\label{fig_flux1}}
\subfigure[mixed][Mixed Composition (adopted from~\cite{Allard:2005cx})]{\includegraphics[width=0.81\linewidth]{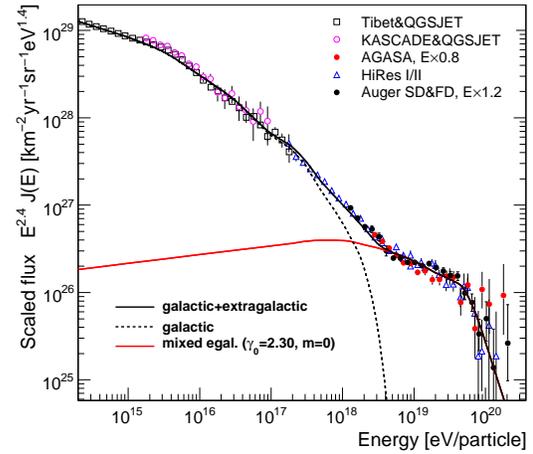}
\label{fig_flux2}}
\subfigure[Transition at the ankle]{\includegraphics[width=0.81\linewidth]{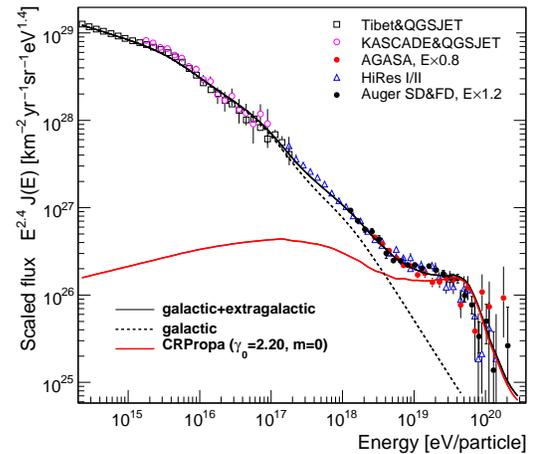}
\label{fig_flux3}}
\end{center}
\caption[transition]{Models of the transition from galactic to
  extra-galactic cosmic rays vs.\ measurements of the all particle
  flux~\cite{Takeda:2002at, Abbasi:2007sv,Yamamoto:2007xj,
    Antoni:2005wq, Tibet_APJ2008}.}
\label{fig_flux}
\end{figure}

In {\itshape mixed composition models}~\cite{Allard:2005cx, Hooper:2006tn} the
extra-galactic cosmic ray composition is assumed to be equal to
the one measured at low energies in our galaxy. Due to the
Lorentz-factor dependence of the energy loss, the individual spectra
of nuclei with mass $A$ are subsequently suppressed at energies above 
$\gtrsim A\cdot 10^{18}$~\eV{}.\\
As in case of the dip-model, this ansatz gives a good description
of the existing data (cf. Fig.~\ref{fig_flux}b),
but with a much softer extra-galactic source spectrum.
The transition from galactic to extra-galactic
cosmic rays is at about a factor 10 higher energies close to 1~\EeV{}
and correspondingly, this model needs galactic sources with a higher maximum
acceleration energy than the dip-model.
\subsection{Ankle Model}
Finally, the traditional way to reproduce the ankle-feature 
is to model it as the intersection of 
a flat extra-galactic component with a 
steep galactic component 
(see for instance~\cite{Wibig:2004ye, Bahcall:2002wi}). In that case,
as can be seen in Fig.~\ref{fig_flux}c), the galactic cosmic ray
spectrum extends to energies well above several~\EeV{}
and thus would require a significant modification of the simple 
rigidity model of the knee.

\section{Composition of UHECRs}
\label{sec:compo}
All of the transition models explained in the last section 
give a similar good description of the measured 
cosmic ray spectra under very different astrophysical assumptions.
Since they differ substantially in the predicted cosmic ray
composition as a function of energy, this observable is the
key to distinguish between the models.

The mass composition estimated from surface detector observables
is compatible with large contributions from heavy elements up
to the highest energies (see~\cite{Dova:2003an,Watson:2004rg}). 
These
estimates rely to a large extent on an accurate prediction of the number
of muons (cf. Eq.~\ref{eq:sizevsE}) in air showers. 
However, modern hadronic interaction models differ
by as much as 30\% in the number of muons on 
ground~\cite{Drescher:2003gh,Pierog:2006qv}. 
Moreover, the application of air shower universality  to data from
the Pierre Auger Observatory suggests, that current 
air shower simulations systematically underestimate
the number of muons~\cite{Schmidt:2007vq,Engel:2007cm}.

\begin{figure*}
\subfigure[Xmax Simulations][Comparison to simulations~\cite{Bergmann:2006yz} 
for proton and iron induced air showers using different hadronic interaction models~\cite{Kalmykov:1997te,Engel:1999db,Ostapchenko:2004ss,Pierog:2006qv}]{\includegraphics[width=0.48\linewidth]{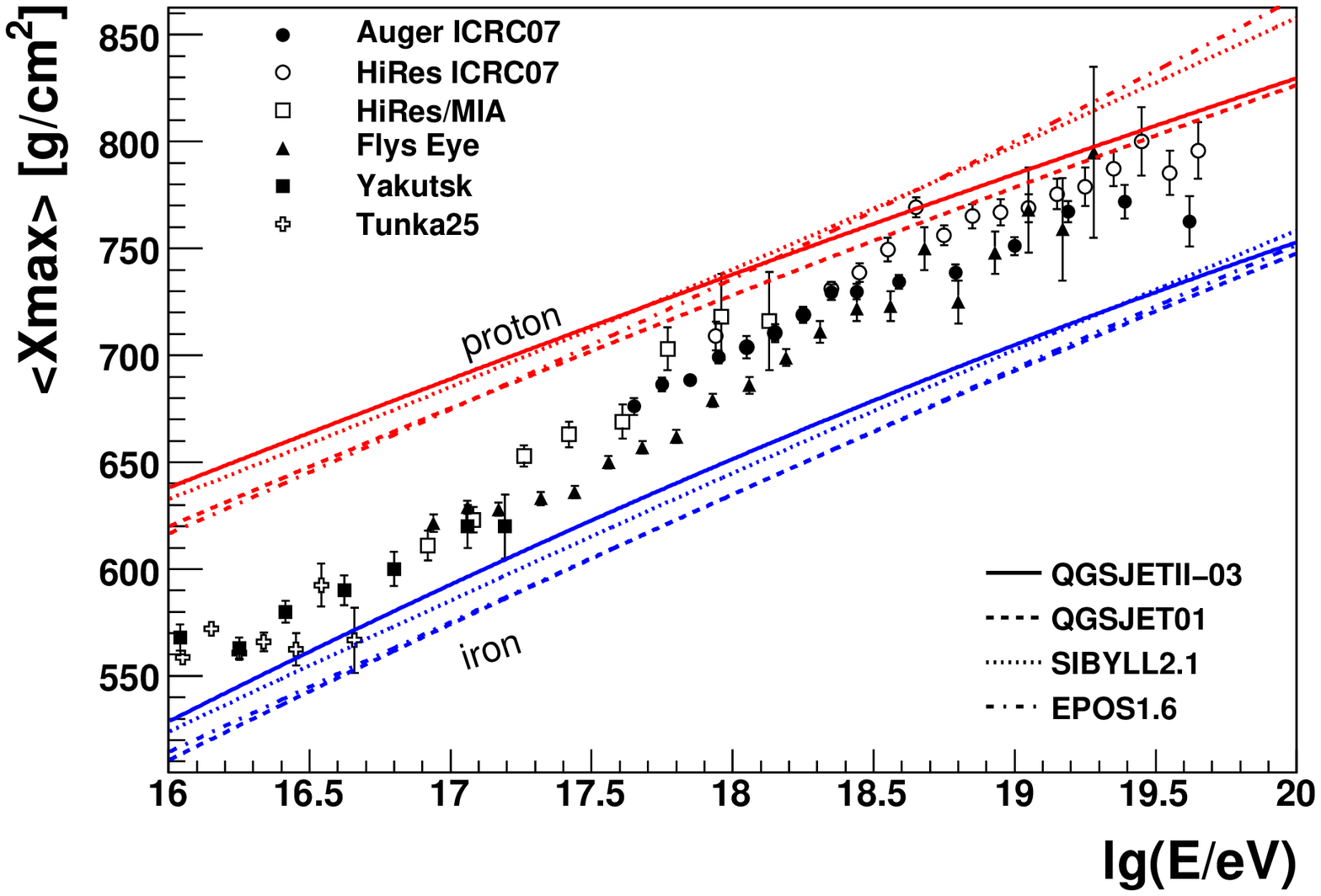}
\label{fig_xmaxSim}}\hfill
\subfigure[Transition models][Comparison to transition models (calculations taken from~\cite{Allard:2007gx}, bands indicate the uncertainty from interaction models).]{\includegraphics[width=0.48\linewidth]{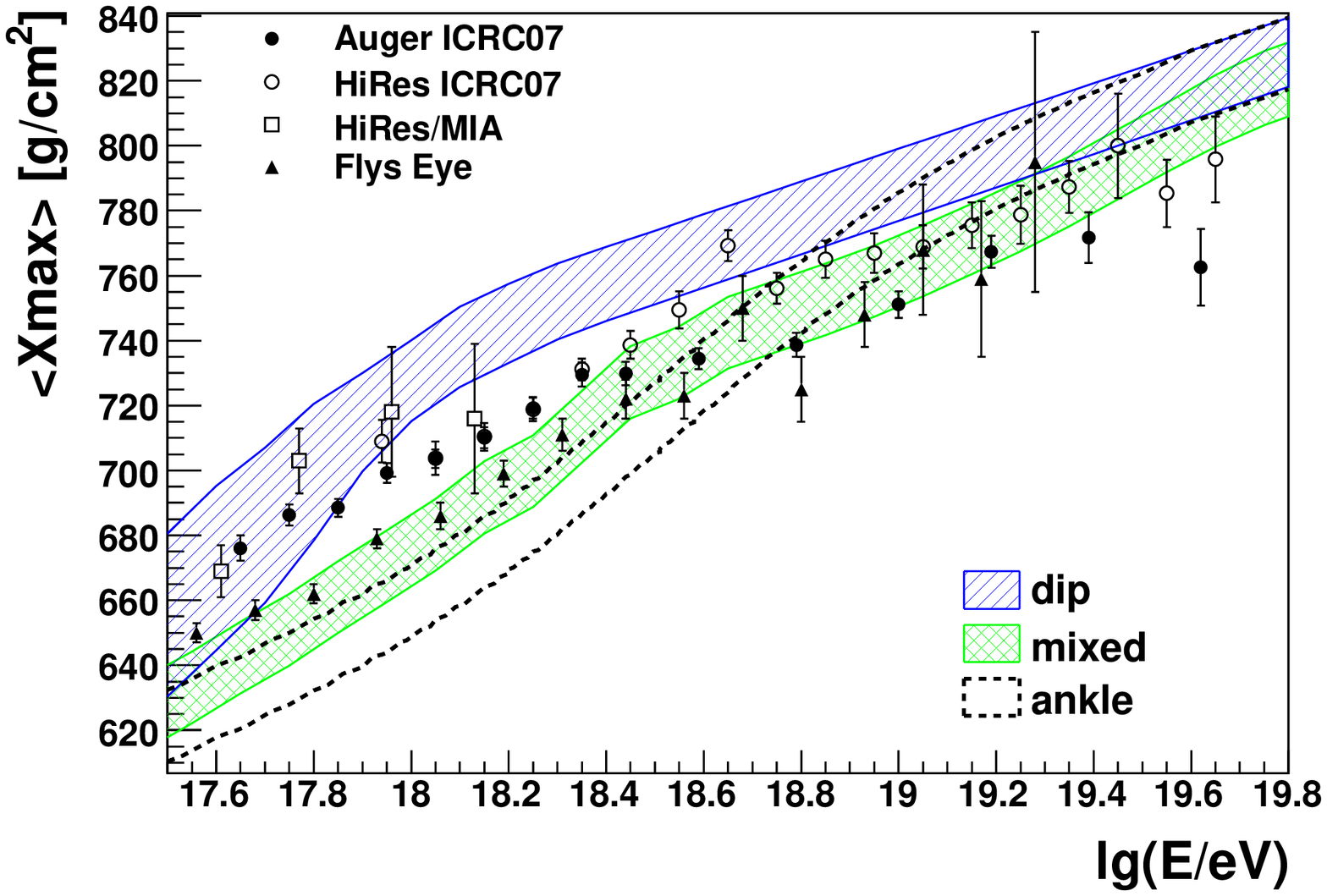}
\label{fig_xmaxTrans}}
\caption[Xmax measurements]{Measurements of $\meanXmax$ from Cherenkov~\cite{knurenko2001,ecrsProsin} and fluorescence~\cite{Bird:1993yi,AbuZayyad:2000ay,Fedorova2007,Unger:2007mc} detectors}
\label{fig_xmax}
\end{figure*}

The maximum of the longitudinal development of the electromagnetic
component of air showers (cf. Eq.~\ref{eq:xmax}) 
provides a composition sensitivity that is somewhat less dependent
on the details of hadronic interactions. As 
can be seen in Fig.~\ref{fig_xmax}a, 
all current air shower models give similar 
predictions of $\meanXmax$ between 0.1 and 10~\EeV{}. It is
however worthwhile noting that this might be a mere coincidence,
since also the predictions of the depth of the shower maximum can be changed
significantly, if some more drastic (though experimentally not
excluded) modifications of the hadronic interactions at high
energies are assumed~\cite{Drescher:2004rc,Ulrich:2007xv,
AlvarezMuniz:2008tu,ecrsWibig}.

The shower maximum can be directly measured by fluorescence detectors,
that can infer the longitudinal shower development from the observation
of fluorescence and Cherenkov light emitted by the shower as a function
of height~\cite{Unger:2008uq}. Observations
of the lateral distribution of Cherenkov light at ground and its pulse 
shape are sensitive to $\Xmax$ as 
well~\cite{Patterson:1983pc,*Patterson:1983qj}.

Measurements of the average of $\Xmax$ over almost three orders of 
magnitude in energy are shown in Fig.~\ref{fig_xmax}a together
with predictions from air shower simulations for proton and iron primaries.
As can be seen, the data are indeed showing a trend from a
heavy composition at low energies towards a light one at
high energies, as would be expected from the transition models
introduced in the last section. There are however systematic differences
between the different experiments. The HiRes data, for instance, 
is compatible with a pure proton composition if compared to the 
{\scshape QGSJET} prediction, whereas the data from the Pierre Auger
Observatory favors a mixed composition at all energies.

A direct comparison of the data to the $\meanXmax$ predicted 
by the dip-, ankle- and mixed-composition model is shown in
Fig.~\ref{fig_xmax}b. Obviously, none of the three models gives 
a satisfactory description of the data, neither in shape nor 
the absolute $\meanXmax$ value, but note that mixed-composition
models have in principle enough parameters to be adjusted to the
data.

Until now we only discussed the average value of the shower maximum.
The {\itshape distribution} of $\Xmax$ can potentially
constrain the mass composition of cosmic rays even better. In the naive
superposition model, one would expect that nuclei with mass $A$ have
smaller shower-to-shower fluctuations by a factor of
$1/\sqrt{A}$. Correctly accounting for nuclear fragmentation leads to
somewhat larger fluctuations of nucleus-induced
showers~\cite{Engel:1992vf, Kalmykov:1993qe,Schatz:1994hv}, but still
the width of the $\Xmax$ distribution of iron showers is about a
factor three smaller than that for proton (about 20 and \ 60~g/cm$^2$
at 1~EeV{} respectively). The analysis of the $\Xmax$ distribution
requires however a good understanding of the detector resolution and
corresponding composition estimates from the $\Xmax$ fluctuations are
still contradictory (for instance pure proton in~\cite{Aloisio:2007rc}
and mixed in~\cite{Knurenko:2004sk} above 1~\EeV{}).\\

\section{The End of the Cosmic Ray Spectrum}
\label{sec:gzk}
\begin{figure}
\includegraphics[width=\linewidth]{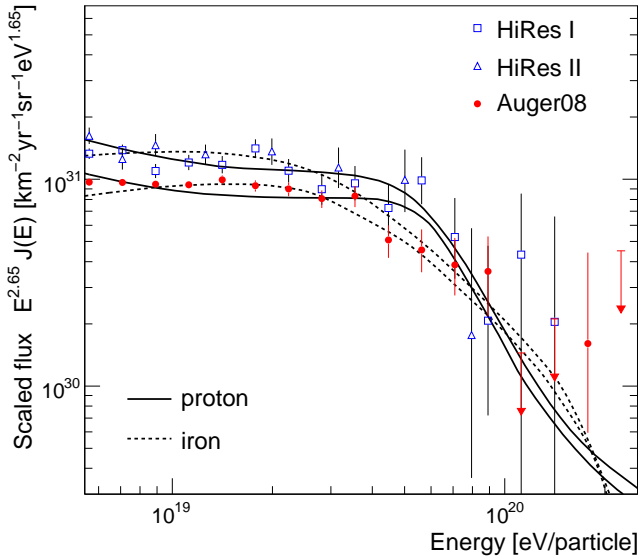}
\caption[gzk]{Measured flux at ultra-high energies compared to
         predictions for propagated proton and iron primaries
         (lines adapted from~\cite{Aloisio:2007rc} and~\cite{Allard:2008gj}).} 
\label{fig_gzk}
\end{figure}

More than forty years after Greisen, Zatsepin and
Kuzmin~\cite{Greisen:1966jv,*Zatsepin:1966jv} (GZK) predicted a suppression
of the cosmic ray flux due to interactions with the cosmic microwave
background (CMB) radiation and its existence has now finally been established
with high significance by HiRes and 
the Pierre Auger Observatory~\cite{Abbasi:2007sv,Abraham:2008ru} 
(cf.~Fig~\ref{fig_gzk}). Furthermore, Auger reported an anisotropy
of the arrival direction of cosmic rays above 
60~\EeV{}~\cite{Cronin:2007zz,*Abraham:2007si,*ecrsSantos} 
and set a limit of $\le 2\%$ on the fraction of photons above 10 EeV{}.
The latter excludes most of the {\itshape top-down} scenarios,
i.e. cosmic ray production in decays of ultra-massive particles
(see eg.~\cite{Kachelriess:2008bk}), that were motivated by the
absence of a GZK feature in the AGASA spectrum~\cite{Takeda:2002at}.\\
The onset of the anisotropy at about the same energy as the GZK threshold
suggests that the suppression is indeed  due to propagation effects
and not because the maximum energy of the sources is reached:
If sources are isotropically distributed on large scales, local anisotropies
can not be detected in a transparent universe, but only if propagation losses
limit the distance from which cosmic rays can reach earth (the so-called
{\itshape GZK-horizon}).

It is a curiosity, that the thresholds for photo-pion production
of protons with photons of the CMB is at a similar energy as the giant dipole
resonance for iron nuclei. As can be seen in Fig.~\ref{fig_gzk}, the 
current statistical precision of the flux measurements at ultra-high energies
is not sufficient to distinguish between the predictions
for the spectral shape for cosmic rays with a pure proton~\cite{Aloisio:2007rc}
and iron~\cite{Allard:2008gj} composition at the source. (It is
 worthwhile noting that the measured spectra are not corrected
for the corresponding experimental energy resolution and if a
deconvolution was applied 'true' shape of the flux suppression 
would get steeper). A possible way to resolve this degeneracy in the near 
future would be the detection of photons~\cite{Risse:2007sd} or
neutrinos~\cite{Engel:2001hd} 
originating from the decay of pions produced during the proton propagation
(nuclei are expected to produce much less 
neutrinos~\cite{Ave:2004uj,Hooper:2004jc, Anchordoqui:2007fi}).

\begin{figure}
\includegraphics[clip,bb=0 -258 1567 1155,width=\linewidth]{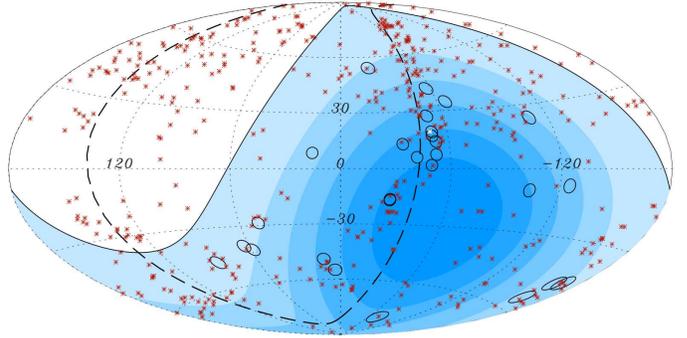}
\caption[agn]{Arrival directions of cosmic rays ($E>60$~\EeV) measured
by the Pierre Auger Observatory (open circles) within its acceptance (shaded
area) and location of active galactic nuclei (small dots)~\cite{Cronin:2007zz,*Abraham:2007si,*ecrsSantos}.}
\label{fig_augerAGN}
\end{figure}

The anisotropy reported by the Pierre Auger Observatory was established by
correlation of the arrival direction of cosmic rays and the location of
active galactic nuclei (AGNs)~\cite{veronCetty} (cf.~Fig.\ref{fig_augerAGN})
within 3.1$^\circ$. This angular scale is compatible
with the deflections expected for protons in the galactic magnetic field
(see for instance \cite{AlvarezMuniz:2001vf}), but from the current
statistics it is not possible to distinguish if the AGNs 
from~\cite{veronCetty} are indeed the source of ultra-high energy
cosmic rays or just a tracer of the true source distribution
like the super-galactic plane~\cite{Stanev:2008sd},
the large scale structure of nearby matter~\cite{Kashti:2008bw} or
even a few sources 
producing nuclei of intermediate mass~\cite{Wibig:2007pf} that
are spread out by magnetic fields. Note that
a follow-up analysis of HiRes did not show a correlation~\cite{Abbasi:2008md},
which may, however, be explained by the different energy scales of the 
two experiments. 

\section{Outlook}
\label{sec:outlook}
The last years have brought a wealth of new precise data on 
cosmic rays above the knee, especially at ultra-high energies
collected by the southern part of the Pierre Auger Observatory. 
Since its construction was just finished this year, one can soon
expect updated results with increased statistics. 
Its northern part is planned to be built in Colorado, USA,
and will increase the exposure of the observatory by a factor
of seven and provide full sky coverage for particle 
astronomy~\cite{Nitz:2007ur}. The space-borne experiments 
TUS~\cite{Abrashkin:2005hb} and JEM-EUSO~\cite{ecrsGorodetzky}
will observe air showers from space and investigate the region
above the GZK cutoff.\\
At intermediate energies, the hybrid Telescope Array~\cite{Kawai:2008zza}
started data taking and will study the region around the ankle
with fluorescence detectors and a scintillator array. Both,
the Pierre Auger Observatory and Telescope Array aim at covering
the transition region of galactic to extragalactic cosmic rays
down to $10^{17}$~\eV{} by using
fluorescence telescopes with enlarged field of views and shielded particle 
counters~\cite{Klages:2007zza,Etchegoyen:2007ai,Thomson2007}. The construction
of the low energy enhancement of the southern Auger site is almost finished and 'first
light' is expected in early 2009.
Finally, the end of the galactic cosmic ray spectrum is currently observed 
by IceTop~\cite{ecrsKlepser} and KASCADE-Grande~\cite{ecrsChiavassa} down
to energies of $10^{16}$~\eV{}.

This new cosmic ray detectors will thus cover more than four orders of
magnitude from the knee up to beyond the GZK cutoff.
A number
of laboratory experiments will provide additional measurements to lower
the systematic uncertainties of the cosmic ray measurements:
Many of the modern detectors use fluorescence detectors to calibrate
their energy scale and the absolute value of the fluorescence yield in air 
is one of the major contributions 
to the current energy scale uncertainties of HiRes and Auger.
It is currently re-measured under various
atmospheric conditions by several groups~\cite{Arqueros:2008cx}.
Furthermore, in order to diminish the uncertainties of the hadronic interaction models
employed to interpret the cosmic ray data, more data from controlled interactions
at accelerators are collected. The 
NA61 experiment~\cite{Abgrall:2008zza} at the Super Proton Synchrotron at CERN 
will measure pion-carbon interactions above 300~\GeV{} that are important
for the last stages of the air shower development~\cite{Meurer:2005dt} and
LHCf~\cite{ecrsMuraki},
TOTEM~\cite{Anelli:2008zza} and CASTOR~\cite{Aslanoglou:2007wv} 
at the Large Hadron Collider will provide data on particle production in the
forward region and the proton-proton cross section at center of 
mass energies corresponding to $10^{17}$~\eV{} in
terms of primary cosmic ray energies.\\

\IEEEtriggeratref{91}
\bibliographystyle{myIEEEtran}
\bibliography{ecrs08}

\end{document}